\documentclass[conference,10pt,letterpaper]{IEEEtran}

\ifCLASSINFOpdf
   \usepackage[pdftex]{graphicx}
\else
\fi

\usepackage{comment}

\usepackage{amsmath}
\usepackage{mathtools}

\usepackage{acronym}

\usepackage{amsfonts}
\usepackage{amsthm}
\usepackage{thmtools}

\usepackage{siunitx}

\usepackage{mleftright}
\usepackage{bm}
\usepackage{bbm}

\usepackage{tikz}
\usepackage{pgfplots}

\usepackage[textsize=tiny, disable]{todonotes} %

\usetikzlibrary{external}
\definecolor{tab101}{RGB}{228,26,28}
\definecolor{tab102}{RGB}{55,126,184}
\definecolor{tab103}{RGB}{77,175,74}
\definecolor{tab104}{RGB}{152,78,163}
\definecolor{tab105}{RGB}{255,127,0}

\usepackage[font=footnotesize, skip=8pt]{caption}
\ifCLASSOPTIONcompsoc
  \usepackage[caption=false,font=normalsize,labelfont=sf,textfont=sf]{subfig}
\else
  \usepackage[caption=false,font=footnotesize]{subfig}
\fi

\newcommand\blfootnote[1]{%
  \begingroup
  \renewcommand\thefootnote{}\footnote{#1}%
  \addtocounter{footnote}{-1}%
  \endgroup
}

\usepackage{graphicx}
\usepackage{color}
\usepackage{pgf, tikz, pgfplots}
\usetikzlibrary{shapes,arrows.meta, calc}
\usetikzlibrary{external}
\usepackage[ruled,vlined, linesnumbered]{algorithm2e}
\let\oldnl\nl%
\newcommand{\nonl}{\renewcommand{\nl}{\let\nl\oldnl}}%

\newcommand\vek[1]{{\bm{#1}}}

\newcommand\ebno{E_{\text{b}}/N_0}

\DeclareMathOperator*{\maxstar}{max^\star}

\definecolor{KITgreen}{rgb}{0,.59,.51}
\definecolor{KITpalegreen}{RGB}{130,190,60} 
\definecolor{KITblack}{rgb}{0,0,0}
\definecolor{KITblue}{rgb}{.27,.39,.66}
\definecolor{KITred}{rgb}{.63,.13,.13}
\definecolor{KITpurple}{rgb}{.64,.06,.48}
\definecolor{KITcyan}{rgb}{.14,.63,.87}
\definecolor{KITyellow}{rgb}{.98,.89,0}
\definecolor{KITorange}{rgb}{.87,.60,.10}

    \acrodef{AIR}[AIR]{achievable information rate}
    \acrodef{APP}[APP]{a posteriori probability}
    \acrodefplural{APP}[APPs]{a posteriori probabilities}
    \acrodef{AWGN}[AWGN]{additive white Gaussian noise}
    \acrodef{BER}[BER]{bit error rate}
    \acrodef{BICM}[BICM]{bit-interleaved coded modulation}
    \acrodef{BMD}[BMD]{bit-metric decoder}
    \acrodef{BMI}[BMI]{bitwise mutual information}
    \acrodef{BP}[BP]{belief propagation}
    \acrodef{BPSK}[BPSK]{binary phase-shift keying}
    \acrodef{BSC}[BSC]{binary symmetric channel}
    \acrodef{CEL}[CEL]{Communications Engineering Lab}
    \acrodef{CM}[CM]{Coded Modulation}
    \acrodef{CPU}[CPU]{central processing unit}
    \acrodef{DFE}[DFE]{decision-feedback equalization}
    \acrodef{DL}[DL]{deep learning}
    \acrodef{DNN}[DNN]{deep neural network}
    \acrodef{FN}[FN]{factor node}
    \acrodef{FFG}[FFG]{Forney-style factor graph}
    \acrodef{FEC}[FEC]{forward error correction}
    \acrodef{FIR}[FIR]{finite impulse response}
    \acrodef{GMI}[GMI]{generalized mutual information}
    \acrodef{GPU}[GPU]{graphics processing unit}
    \acrodef{GNN}[GNN]{graph neural network}
    \acrodef{iid}[i.i.d.]{independent and identically distributed}
    \acrodef{ISI}[ISI]{inter-symbol interference}
    \acrodef{KIT}[KIT]{Karlsruhe Institute of Technology}
    \acrodef{LDPC}[LDPC]{low-density parity-check}
    \acrodef{LLR}[LLR]{log-likelihood ratio}
    \acrodef{LR}[LR]{likelihood ratio}
    \acrodef{MAP}[MAP]{maximum a posteriori}
    \acrodef{MIMO}[MIMO]{multiple-input multiple-output}
    \acrodef{ML}[ML]{maximum likelihood}
    \acrodef{MLSE}[MLSE]{maximum likelihood sequence estimation}
    \acrodef{MMSE}[MMSE]{minimum mean squared error}
    \acrodef{MSE}[MSE]{mean squared error}
    \acrodef{NBP}[NBP]{neural belief propagation}
    \acrodef{NN}[NN]{neural network}
    \acrodef{pdf}[PDF]{probability density function}
    \acrodef{pmf}[PMF]{probability mass function}
    \acrodef{QAM}[QAM]{quadrature amplitude modulation}
    \acrodef{RV}[RV]{random variable}
    \acrodef{SER}[SER]{symbol error rate}
    \acrodef{SISO}[SISO]{soft-input soft-output}
    \acrodef{SNR}[SNR]{signal-to-noise ratio}
    \acrodef{SPA}[SPA]{sum-product algorithm}
    \acrodef{SPC}[SPC]{single parity check}
    \acrodef{TEQ}[TEQ]{turbo-equalization}
    \acrodef{VN}[VN]{variable node}

\pgfplotsset{compat=newest}
\begin{document}
\title{Neural Enhancement of Factor Graph-based \\ Symbol Detection\vspace*{-0.5ex}}

\author{\IEEEauthorblockN{Luca Schmid and Laurent Schmalen}
\IEEEauthorblockA{Communications Engineering Lab (CEL), Karlsruhe Institute of Technology (KIT)\\ 
Hertzstr. 16, 76187 Karlsruhe, Germany, 
Email: \texttt{\{first.last\}@kit.edu}\vspace*{-1ex}}
}
\markboth{Submitted version, \today}%
{Schmid \MakeLowercase{\textit{et al.}}: XXX}

\maketitle
\begin{abstract}
We study the application of the factor graph framework for symbol detection on linear inter-symbol interference channels.
Cyclic factor graphs have the potential to yield low-complexity symbol detectors, but are suboptimal if the ubiquitous sum-product algorithm is applied.
In this paper, we present and evaluate strategies to improve the performance of cyclic factor graph-based symbol detection algorithms by means of neural enhancement. In particular, we apply neural belief propagation as an effective way to counteract the effect of cycles within the factor graph. We further propose the application and optimization of a linear preprocessor of the channel output. By modifying the observation model, the preprocessing can effectively change the underlying factor graph, thereby significantly improving the detection performance as well as reducing the complexity.
\blfootnote{This work has received funding in part from the European Research Council (ERC) under the European Union’s Horizon 2020 research and innovation programme (grant agreement No. 101001899) and in part from the German Federal Ministry of Education and Research (BMBF) within the project Open6GHub (grant agreement 16KISK010).}
\end{abstract}

\IEEEpeerreviewmaketitle
\section{Introduction}
Factor graphs are a flexible tool for
algorithmic modeling of efficient inference algorithms.
By representing the factorization of a composite global function of many variables in a graphical way, the computation of various marginalizations of this function can be efficiently implemented by a message passing algorithm.
That is why many well-known algorithms like the Viterbi algorithm, the BCJR algorithm or Pearl's belief propagation can be derived by message passing on a suitable factor graph \cite{kschischang_factor_2001}.
While being originally derived for tree-structured graphs, message passing schemes, like
the \ac{SPA}, can also be applied to cyclic graphs, which leads to iterative, suboptimal algorithms.

In this paper, we investigate the behavior of the \ac{SPA} on cyclic factor graphs for the well-known task of a transmission over an \ac{AWGN} channel affected by linear \ac{ISI}.
Modeling a factor graph based on the \emph{Forney} observation model~\cite{forney_lower_1972} leads to a detection algorithm with a complexity that is linear in the block length but exponential in the number of interferers~\cite{colavolpe_application_2005}. 
Alternatively, the symbol detector in~\cite{colavolpe_siso_2011} employs a factor graph based on the \emph{Ungerboeck} observation model~\cite{ungerboeck_adaptive_1974} and has significantly reduced complexity. However, both algorithms show a greatly differing performance and the low-complexity detector may suffer from large performance penalties.

Recently, model-based deep learning has shown great potential to empower various communication algorithms~\cite{shlezinger_model-based_2020} and overcome their limitations.
In~\cite{shlezinger_inference_2020}, the \acp{FN} of a cycle-free factor graph are replaced by \acp{DNN} that are utilized to learn the local mappings of the \acp{FN},
thereby robustifying the algorithm towards model uncertainties.
To mitigate the performance loss for cyclic factor graphs,
 a \ac{GNN}, which is structurally identical to the original graph, but has fully parametrized message updates is used in~\cite{satorras_neural_2021}.
The \ac{GNN} runs conjointly to the original algorithm
and corrects the \ac{SPA} messages after each iteration.
The authors in \cite{liu_novel_2021} compensate the performance degradation due to cycles in the graph by concatenating a supplemental \ac{NN}-based \ac{FN} to the factor graph. This additional \ac{FN} is connected to all \acp{VN} and is optimized in and end-to-end manner. However, the underlying \ac{NN} structure is specifically tailored to binary transmission which substantially limits its scope of application.

Instead of replacing different components of the factor graph by \acp{DNN},
 the \ac{SPA} is unfolded to a \ac{DNN} and the resulting graph is equipped with tunable weights in~\cite{nachmani_learning_2016}. This approach is known as \ac{NBP}.
In this paper, we consider \ac{NBP} for the application of factor graph-based symbol detection.
Since the performance improvement of \ac{NBP} is limited, 
we propose a novel generalization of the factor graph by introducing additional multiplicative weights within the \acp{FN}.
Optimizing all weights in an end-to-end manner already leads to considerable performance gains.
Moreover, we leverage the high sensibility of the \ac{SPA} to a variation of the underlying graph structure by applying an optimizable linear filter to the channel output, which allows us to modify the observation model and thereby the factor graph itself.
Varying the filter order, the proposed generalization provides an adjustable performance-complexity tradeoff and can significantly outperform conventional factor graph models, which we illustrate by simulations for two different \ac{ISI} channels and two different modulation formats.

\section{Factor Graphs and Marginalization}\label{sec:factor_graph}

Let $f(\mathcal{X})$ be a multivariate function which depends on the set of variables ${\mathcal{X} = \{ x_0, \ldots x_n \}}$
and which can be factorized as
\begin{equation} \label{eq:factorization}
    f(\mathcal{X}) = \prod\limits_{j=1}^J f_j(\mathcal{X}_j), \qquad \mathcal{X}_j \subset \mathcal{X}.
\end{equation}
A \emph{factor graph} represents the factorization of $f(\mathcal{X})$
in a graphical way~\cite{kschischang_factor_2001}. 
The following rules define the bijective relationship between \eqref{eq:factorization} and its corresponding factor graph:
\begin{itemize}
    \item Every factor $f_j(\mathcal{X}_j)$ is represented by a unique vertex, the so-called \ac{FN} $f_j$.
    \item Every variable $x \in \mathcal{X}$ is represented by a unique vertex, the so-called
    \ac{VN} $x$.
    \item An \ac{FN} $f_j$ is connected to a \ac{VN} $x$
          if and only if the corresponding factor $f_j(\mathcal{X}_j)$ is a function of $x$, 
          i.e., if ${x \in \mathcal{X}_j =: \mathcal{N}(f_j)}$.
\end{itemize}
The notation $\mathcal{N}(f_j)$ is called the \emph{neighborhood} of the \ac{FN} $f_j$.
Equivalently, ${\mathcal{N}(x) := \{ f_j : \, x \in \mathcal{X}_j, j=1,\ldots,J \}}$ denotes the neighborhood of the \ac{VN} $x$.

The \ac{SPA} is a message passing algorithm which computes
the marginalization $f(x_i)$ of the global function $f(\mathcal{X})$ with respect to each variable
${x_i \in \mathcal{X}}$, respectively. By operating on a factor graph, it implicitly
leverages the distributive law on the factorization of $f(\mathcal{X})$.
Messages are propagated between the nodes of the factor graph along its edges 
and represent interim results of the marginalization. 
Let $\mu_{f_j \rightarrow x}(x)$ denote a message sent 
from \ac{FN} $f_j$ along an edge to \ac{VN} $x$.
Consequently, $\mu_{x \rightarrow f_j}(x)$ denotes  a message on the same edge, but sent in the
opposite direction.
The message passing algorithm is based on one central message update rule for the \acp{FN} and
\acp{VN}, respectively.
In the logarithmic domain,
the message updates are
\begin{align}
\mu_{x \rightarrow f_j}(x)
    &= \sum\limits_{f' \in \mathcal{N}(x) \setminus \{ f_j \}} \mu_{f' \rightarrow x} (x) 
    \label{eq:v2f_update}\\
    \mu_{f_j \rightarrow x}(x)
    &= \maxstar\limits_{\sim \{ x \}}
    \mleft( \ln \mleft( f_j(\mathcal{X}_j) \mright) + \sum\limits_{\mathclap{x' \in \mathcal{N}(f_j) \setminus \{ x \}}} \mu_{x' \rightarrow f_j} (x') \mright). \label{eq:f2v_update}
\end{align}
The operator $\maxstar \delta_i$ denotes the Jacobian logarithm \cite{robertson_comparison_1995} which computes ${\ln\mleft(\mathrm{e}^{\delta_1} + \ldots + \mathrm{e}^{\delta_n} \mright)}$.
Applying the Jacobian logarithm over all variables of a function except $x$ is denoted by the \emph{summary} operator $\maxstar_{\sim \{x\}}$.
Based on the \ac{SPA} message update rule, we can compute marginals by propagating messages through the respective factor graph.
If the graph is tree-structured, messages travel forward and backward through the entire graph, starting at the leaf nodes.
Based on the computed messages, the exact marginals
\begin{equation}\label{eq:final_marginal}
    f(x_i) = \exp \mleft( \sum\limits_{f' \in \mathcal{N}(x_i)} \mu_{f' \rightarrow x_i}(x_i) \mright), \qquad x_i \in \mathcal{X}
\end{equation}
can be obtained.
Since the message updates are local~\cite{kschischang_factor_2001}
and because the \ac{SPA} makes no reference to the topology of the graph~\cite{yedidia_understanding_2003},
the \ac{SPA} may also be applied to factor graphs with cycles, yielding an iterative algorithm.
The messages are initialized with an unbiased state in iteration
${n=0}$ and are iteratively updated by following a certain schedule 
until convergence or a stopping criterion is reached.
In the case of cyclic factor graphs, the superscript $(n)$, with ${n=0,\ldots,N}$ indicates the iteration in which the message $\mu_{a \rightarrow b}^{(n)}$ is computed.
In contrast to tree-structured graphs, this iterative algorithm 
only yields an approximation of the exact marginals~\cite{kschischang_factor_2001}.
However, many successful applications, e.g., decoders of error-correcting codes, are based on message passing on cyclic graphs~\cite{yedidia_understanding_2003}.

\section{Symbol Detection}\label{sec:channel_model}
We consider the transmission of an information sequence 
${\vek{c} \in \mathcal{M}^K}$ of a multilevel
constellation $\mathcal{M} = \{\text{m}_i \in \mathbb{C}, i=1,\ldots,M \} $ 
over a baseband channel, impaired by linear inter-symbol interference and \ac{AWGN}.
The bit pattern of length $m :=\log_2 \mleft( M \mright)$ which corresponds to a
symbol $c_k$ is denoted by ${\vek{b}(c_k)} = \left( b_1(c_k),\ldots,b_m(c_k) \right)$.
The relationship between the independent and uniformly distributed symbols $c_k$ and the receive symbols $y_k$ can be expressed by an equivalent discrete-time channel model~\cite{forney_lower_1972}:
\begin{equation}\label{eq:convolution}
    y_k = \sum\limits_{\ell=0}^{L} h_{\ell} c_{k-\ell} + w_k,
    \qquad k = 1,\ldots,K\!+\!L.
\end{equation}
For a channel with memory $L$, ${\vek{h} \in \mathbb{C}^{L+1}}$ is the finite channel impulse response and $\vek{w} \sim \mathcal{C}\mathcal{N}(0,\sigma^2 \boldsymbol{I})$ 
denotes white circular Gaussian noise.
We assume that the symbols $c_k$ for ${k\leq1}$ and ${k > K}$ represent the known initial state and the steady state of the channel filter.
Since the interference is linear, \eqref{eq:convolution} can
be described in matrix vector notation
\begin{equation*}
    \vek{y} = \vek{H} \check{\vek{c}} + \vek{w}, \quad \vek{H} \in \mathbb{C}^{(K+L) \times (K+2L)}
\end{equation*}
with the equivalent transmit sequence $\check{\vek{c}} := (c_{1-L},\ldots,c_{K+L})$.

\begin{figure}[tb]
    \centering
    \tikzstyle{fn} = [draw, very thick, regular polygon, regular polygon sides=4, minimum width = 2.5em, inner sep=0pt]
\tikzstyle{fn_ghost} = [very thick, regular polygon, regular polygon sides=4, minimum width = 2.5em, inner sep=0pt]
\tikzstyle{vn} = [draw, very thick, circle, inner sep=0pt, minimum size = 2em]
\tikzstyle{vn_ghost} = [very thick, circle, inner sep=0pt, minimum size = 2em]

\begin{tikzpicture}[auto, node distance=3em and 3em, thick]
\clip (1.5, -0.37) rectangle (9.2, 2.15);
    \node [vn_ghost] (c0){};
    \node [vn_ghost, right= of c0] (c1){$\cdots$};
    \node [vn, label=center:$c_{2}$, right= of c1] (c2){};
    \node [vn, label=center:$c_{3}$, right= of c2] (c3){};
    \node [vn, label=center:$c_{4}$, right= of c3] (c4){};
    \node [vn_ghost, right= of c4] (c5){$\cdots$};
    \node [vn_ghost, right= of c5] (c6){};
    
    \node [fn_ghost, above= of c0] (f0){};
    \node [fn_ghost, above=of c1] (f1){};
    \node [fill=KITblue, fill opacity=0.2, fn, label=center: $q_{2}$, above= of c2] (f2){};
    \node [fill=KITblue, fill opacity=0.2, fn, label=center: $q_{3}$, above= of c3] (f3){};
    \node [fill=KITblue, fill opacity=0.2, fn, label=center: $q_{4}$, above= of c4] (f4){};
    \node [fn_ghost, above= of c5] (f5){};
    \node [fn_ghost, above= of c6] (f6){};
    
    \draw[-] (f2.south) -- ($(f2.south)!0.3!(c0.north)$);
    \draw[dotted] (f2.south) -- ($(f2.south)!0.5!(c0.north)$);
    \draw[dotted] (f2.south) -- (c1.north);
    \draw[-] (f2.south) -- ($(f2.south)!0.5!(c1.north)$);
    \draw[-] (f2.south) -- (c2.north);
    
    \draw[-] (f3.south) -- ($(f3.south)!0.7!(c1.north)$);
    \draw[dotted] (f3.south) -- (c1.north);
    \draw[-] (f3.south) -- (c2.north);
    \draw[-] (f3.south) -- (c3.north);
    
    \draw[-] (f4.south) -- (c2.north);
    \draw[-] (f4.south) -- (c3.north);
    \draw[-] (f4.south) -- (c4.north);
    
    \draw[dotted] (f5.south) -- (c3.north);
    \draw[-] (c3.north) -- ($(c3.north)!0.7!(f5.south)$);
    \draw[dotted] (f5.south) -- (c4.north);
    \draw[-] (c4.north) -- ($(c4.north)!0.5!(f5.south)$);
    
    \draw[dotted] (c4.north) -- ($(c4.north)!0.5!(f6.south)$);
    \draw[-] (c4.north) -- ($(c4.north)!0.3!(f6.south)$);
\end{tikzpicture}
    \caption{Factor graph representation of \eqref{eq:fgeq} for $L=2$.}
    \label{fig:fgeq}
    \vspace{-8pt}
\end{figure}
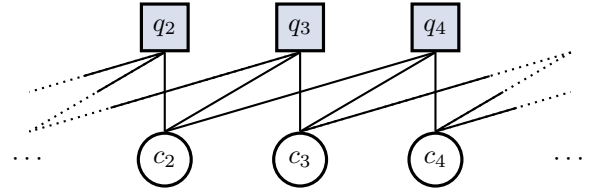
\begin{figure}[tb]
    \centering
    \tikzstyle{fn} = [draw, very thick, regular polygon, regular polygon sides=4, minimum width = 2.5em, inner sep=0pt]
\tikzstyle{fn_ghost} = [very thick, regular polygon, regular polygon sides=4, minimum width = 2.5em, inner sep=0pt]
\tikzstyle{vn} = [draw, very thick, circle, inner sep=0pt, minimum size = 2em]
\tikzstyle{vn_ghost} = [very thick, circle, inner sep=0pt, minimum size = 2em]

\begin{tikzpicture}[auto, node distance=3em and 3.5em, thick]
\clip (1.7, -2.1) rectangle (9.5, 0.4);
    \node [vn_ghost] (c0){};
    \node [vn_ghost, right= of c0] (c1){$\cdots$};
    \node [vn, label=center:$c_{2}$, right= of c1] (c2){};
    \node [vn, label=center:$c_{3}$, right= of c2] (c3){};
    \node [vn, label=center:$c_{4}$, right= of c3] (c4){};
    \node [vn_ghost, right= of c4] (c5){};
    \node [fn_ghost, right=0.5em of c4] (rdots) {$\cdots$};
    \node [vn_ghost, right= of c5] (c6){};

    \node [fill=KITgreen, fill opacity=0.2, fn, label=center: $F_{2}$, left=0.5em of c2] (p2){};
    \node [fill=KITgreen, fill opacity=0.2, fn, label=center: $F_{3}$, left=0.5em of c3] (p3){};
    \node [fill=KITgreen, fill opacity=0.2, fn, label=center: $F_{4}$, left=0.5em of c4] (p4){};

    \node [fill=KITred, fill opacity=0.2, fn, label=center: $I_{2,0}$, below= of c2.260, anchor = north east] (I20){};
    \node [fill=KITred, fill opacity=0.2, fn, label=center: $I_{2,1}$, below= of c2.280, anchor = north west] (I21){};
    \node [fill=KITred, fill opacity=0.2, fn, label=center: $I_{3,1}$, below= of c3.260, anchor = north east] (I31){};
    \node [fill=KITred, fill opacity=0.2, fn, label=center: $I_{3,2}$, below= of c3.280, anchor = north west] (I32){};
    \node [fill=KITred, fill opacity=0.2, fn, label=center: $I_{4,2}$, below= of c4.260, anchor = north east] (I42){};
    \node [fill=KITred, fill opacity=0.2, fn, label=center: $I_{4,3}$, below= of c4.280, anchor = north west] (I43){};
    \node [fn_ghost, below= of c5.260, anchor = north east] (I53){};
    \node [fn_ghost, below= of c5.280, anchor = north west] (I54){};
    \node [fn_ghost, below= of c6.260, anchor = north east] (I64){};
    \node [fn_ghost, below= of c6.280, anchor = north west] (I65){};
    
    \draw[-] (c2.west) -- (p2.east);
    \draw[-] (c3.west) -- (p3.east);
    \draw[-] (c4.west) -- (p4.east);
    \draw[-] (I20.north) -- (c2.south);
    \draw[dotted] (I20.north) -- ($(I20.north)!0.4!(c0.south)$);
    \draw[-] (I20.north) -- ($(I20.north)!0.2!(c0.south)$);
    \draw[-] (I21.north) -- (c2.south);
    \draw[-] (I21.north) -- ($(I21.north)!0.58!(c1.south)$);
    \draw[dotted] (I21.north) -- ($(I21.north)!1.0!(c1.south)$);
   \draw[-] (c2.south) -- (I32.north);
    \draw[-] (c2.south) -- (I42.north);
    \draw[-] (I31.north) -- (c3.south);
    \draw[-] (I31.north) -- ($(I31.north)!0.7!(c1.south)$);
    \draw[dotted] (I31.north) -- ($(I31.north)!1.0!(c1.south)$);
    \draw[-] (I32.north) -- (c3.south);
    \draw[-] (c3.south) -- (I43.north);
    \draw[-] (c3.south) -- ($(c3.south)!0.8!(I53.north)$);
    \draw[dotted] (c3.south) -- ($(c3.south)!1.0!(I53.north)$);
    \draw[-] (I42.north) -- (c4.south);
    \draw[-] (I43.north) -- (c4.south);
    \draw[-] (c4.south) -- ($(c4.south)!0.4!(I54.north)$);
    \draw[dotted] (c4.south) -- ($(c4.south)!0.8!(I54.north)$);
    \draw[-] (c4.south) -- ($(c4.south)!0.28!(I64.north)$);
    \draw[dotted] (c4.south) -- ($(c4.south)!0.4!(I64.north)$);

\end{tikzpicture}
    \caption{Factor graph representation of \eqref{eq:ubfgeq} for $L=2$.}
    \label{fig:ubfgeq}
    \vspace{-8pt}
\end{figure}
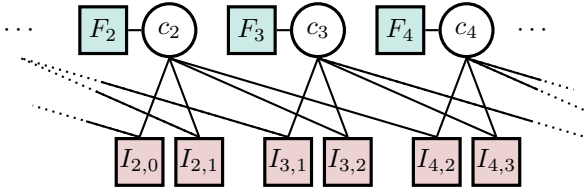

We study the problem of symbol detection, i.e., the estimation of the symbols $c_k$ from the observed sequence $\vek{y}$. 
In the context of Bayesian inference, 
we are interested in the symbol-wise \acp{APP}
\begin{equation}\label{eq:symbol_wise_app}
    P(c_k = \text{m} |\vek{y}) = 
    \sum_{\mathclap{\vek{c} \in \mathcal{M}^K,\, c_k = \text{m}}} P(\vek{c|y}), \qquad k=1,\ldots,K.
\end{equation}
The marginalization in~\eqref{eq:symbol_wise_app} can be efficiently computed by the \ac{SPA} if the \ac{APP} distribution 
$P(\vek{c}|\vek{y})$ can be represented by a suitable factor graph.
Applying Bayes' theorem and exploiting the uniform distribution of the independent information symbols, we obtain
\begin{equation}\label{eq:likelihood}
P(\vek{c}|\vek{y}) \propto p(\vek{y}|\vek{c}) = \frac{1}{(\pi \sigma^2)^{K}} \exp \left ( -\frac{\Vert \vek{y-Hc} \Vert ^2}{\sigma^2} \right ).
\end{equation}
Using the chain rule of conditional probabilities and exploiting the Markovian property of the channel with finite memory $L$ and \ac{AWGN},
we factorize the likelihood function
\begin{align}
    p(\vek{y}|\vek{c}) \propto{}& \prod\limits_{k=1}^{K} \exp \mleft(-{{\vert y_{k}-\sum_{\ell =0}^{L}h_{\ell}c_{k-\ell} \vert^{2}}\over {\sigma^{2}}}\mright) \nonumber \\
    =:{}& \prod\limits_{k=1}^{K} q_k(c_k,c_{k-1}\ldots,c_{k-L})
    \label{eq:fgeq}
\end{align}
and represent it by a factor graph, as illustrated in Fig.~\ref{fig:fgeq}.
By running the \ac{SPA} on this factor graph, we obtain a symbol detection algorithm that we call FFG (Forney-based factor graph symbol detector) and which was initially proposed in~\cite{colavolpe_application_2005}.
Alternatively, we can rewrite the exponential term in \eqref{eq:likelihood} as
\begin{equation*}
    p(\vek{y|c}) \propto \exp\mleft( {2{\text{Re}}\mleft\{ \vek{c}^{\text H} \vek{H}^{\text H} \vek{y} \mright\} -\vek{c}^{\text H}\vek{H}^{\text H}\vek{H}\vek{c}\over \sigma^2}  \mright),
\end{equation*}
where proportionality $\propto$ denotes two terms only differing in a factor
independent of $\vek{c}$.
By substituting
\begin{equation}
    \vek{G} := \vek{H}^{\text H} \vek{H}, \qquad
    \vek{x} := \vek{H}^{\text H} \vek{y} \label{eq:x_matched_filter},
\end{equation}
and using the scalar notation
\begin{align*}
    \vek{c}^{\text H}\vek{x} &= \sum\limits_{k=1}^{K} x_k c_k^{\star} \\
    \vek{c}^{\text H}\vek{G}\vek{c} &= \sum\limits_{k=1}^{K} G_{k,k} |c_{k}|^2 - \sum\limits_{k=1}^{K} \sum\limits_{\substack{\ell=1 \\ \ell \neq k}}^{K} \text{Re} \mleft\{ G_{k,\ell} c_{\ell} c_{k} \mright\},
\end{align*}
an alternative factorization 
\begin{equation}
    p(\vek{y|c}) \propto \prod\limits_{k=1}^{K} \left [ {F_k(c_k) }
        \prod\limits_{\substack{\ell=1 \\ \ell \neq k}}^K  {J_{k,\ell}(c_k,c_\ell) } 
        \right ] \label{eq:ubfgeq}
\end{equation}
with the factors
\begin{align}
    F_k(c_k) &:=  \exp \mleft( \frac{1}{\sigma^2} {\text{Re}} \mleft\{ 2 x_k c_k^\star - G_{k,k}|c_k|^2 \mright\} \mright) \label{eq:f_fn} \\
    J_{k,l}(c_k,c_{\ell}) &:= \exp \mleft(  -\frac{1}{\sigma^2} {\text{Re}} \mleft\{  G_{k,\ell} c_\ell c_k^\star  \mright\}  \mright) \label{eq:j_fn}
\end{align}
can be found~\cite{colavolpe_siso_2011}.
Figure~\ref{fig:ubfgeq} illustrates the factor graph corresponding to~\eqref{eq:ubfgeq}.
The factors $J_{k,\ell}(c_k, c_\ell)$ and $J_{\ell,k}(c_\ell, c_k)$ depend on the same variables and can be condensed to one factor
$I_{k,\ell}(c_k,c_\ell) := J_{k,\ell}(c_k, c_\ell)  J_{\ell,k}(c_\ell, c_k)$ with $k > l$.
Applying the \ac{SPA} on the factor graph in Figure~\ref{fig:ubfgeq}
yields an alternative symbol detection algorithm which we call UFG (Ungerboeck-based factor graph symbol detector).
It was initially proposed by Colavolpe in~\cite{colavolpe_siso_2011}.

For both  FFG and UFG algorithms, all messages are initialized with
${\mu_0(c_k):=-\ln\mleft(M\mright)}$ and we perform $N$ \ac{SPA} iterations. We apply a flooding schedule, i.e., one iteration comprises the simultaneous update of all messages from \acp{VN} to \acp{FN} in a first step, followed by the update of messages propagating in the opposite direction in a second step. The soft output $\hat{P}(c_k|\vek{y})$ is finally obtained by applying~\eqref{eq:final_marginal} to all \acp{VN}.
Although the factorizations \eqref{eq:fgeq} and \eqref{eq:ubfgeq} are both exact, the  FFG and UFG algorithms are both suboptimal due to cycles within the underlying graphs and only yield an approximation for the symbol-wise \acp{APP}.

The complexity of factor graph-based algorithms can be estimated by the number of message updates at the \acp{FN}, since the \ac{FN} update rule~\eqref{eq:f2v_update} is computationally more demanding than the operation~\eqref{eq:v2f_update} at the \acp{VN}.
The asymptotic complexity of the FFG algorithm is $\mathcal{O}(NKLM^L)$~\cite{colavolpe_application_2005}.
In contrast, the UFG algorithm has a significantly reduced complexity in the order of $\mathcal{O}(NKLM^2)$ which only grows linearly with both the block length $K$ and the channel memory $L$~\cite{colavolpe_siso_2011}.
\section{Neural Belief Propagation on Factor Graphs}\label{sec:nbp}
By the use of deep unfolding, first introduced in~\cite{gregor_learning_2010},
an iterative algorithm can be converted into a \ac{DNN}.
Unfolding the $N$ \ac{SPA} iterations on a factor graph is natural since each iteration is already (factor) graph-based.
The resulting unrolled \ac{DNN} comprises $N$ layers and messages are propagated in a feed-forward fashion through the layers of the network.
\acp{VN} and \acp{FN} accept incoming messages from the previous layer and apply the \ac{SPA} message update rule. The resulting outgoing messages are then forwarded downstream to the next layer. As a consequence, each transmitted message in every iteration has its individually assigned edge. By accordingly weighting each message per edge, we attempt to mitigate the effects of short cycles and improve the performance of message passing on cyclic factor graphs. 
Optimizing the weights towards a loss function
by the use of established deep learning techniques is known as \ac{NBP}~\cite{nachmani_learning_2016}.
In consistency with the message notation, the 
message $\mu_{x \rightarrow f}^{(n)}(x)$ is multiplied by the 
weight $w_{x \rightarrow f}^{(n)}$.
We parametrize the \ac{NBP} of the FFG and UFG algorithms by
\begingroup\abovedisplayskip=8pt\belowdisplayskip=0pt
\begin{align*}
    \mathcal{P}_\text{FFG} := \{ &w_{p_k \rightarrow c_{k-\ell}}^{(n)}, w_{c_{k-\ell} \rightarrow p_k}^{(n)},\,\ell=0,\ldots,L, \nonumber \\
    &n=1\ldots,N,\, k=1\ldots,K\!+\!L\},
\end{align*}
\endgroup
\begingroup\abovedisplayskip=2pt\belowdisplayskip=8pt
and
\begin{align*}
    \mathcal{P}_\text{UFG}(L) :=  \Big\{ &w_{c_k \rightarrow I_{k,\ell}}^{(n)}, w_{I_{k,\ell} \rightarrow c_k}^{(n)}, \, n = 1,\ldots, N,  \nonumber \\ 
    & k = 1,\ldots, K\!+\!L, \, |k-\ell|=1,\ldots,L \Big\}.
\end{align*}
\endgroup
The \acp{FN} $F_k$ have degree~$1$. Incident edges are thus not included in the cycles of the factor graph and are consequently not weighted.
The computational complexity of \ac{NBP} is slightly increased compared to the \ac{SPA}. $M$ additional multiplications per message are required.

The UFG algorithm can be further generalized by varying the observation model on which the factor graph is based.
The original factorization in \eqref{eq:x_matched_filter} implicitly uses a preprocessor by applying a matched filter $\vek{P} = \vek{H}^{\text H}$  to the channel observation $\vek{y}$. 
Inference is carried out using the new observation $\vek{x}$, based
on the so called Ungerboeck observation model~\cite{ungerboeck_adaptive_1974} which is characterized by
$\vek{G}$ and directly affects the underlying factor graph in \eqref{eq:f_fn} and \eqref{eq:j_fn}.
We generalize the preprocessor to a generic \ac{FIR} filter by modifying the observation model to
\begin{equation*}
    \vek{\tilde{G}} := \vek{P}\vek{H},\qquad
    \vek{\tilde{x}} := \vek{P}\vek{y}, 
\end{equation*}
where $\vek{P}$ is a convolutional Toeplitz matrix based on the generic impulse response $\vek{p}\in \mathbb{R}^{L_\text{p}}$ of an \ac{FIR} preprocessing filter.
Applying the \ac{SPA} on this modified factor graph yields a novel detection algorithm which we call $\text{GFG}(\vek{P})$.
The instance $\text{GFG}(\vek{H}^{\text H})$ is equivalent to the UFG equalizer.
Additionally, we generalize the \acp{FN} of the $\text{GFG}$ algorithm by the application of multiplicative weights $\kappa_{i,k}^{(n)}$ and $\lambda_{k,\ell}^{(n)}$ in order to increase the parameter optimization space further.
The resulting factors, given in the logarithmic domain, are
\begingroup\abovedisplayskip=2pt
\begin{align*}
    \tilde{F}^{(n)}_k(c_k) &:=\frac{\kappa^{(n)}_{1,k}}{\sigma^2} {\text{Re}} \mleft\{ {\kappa^{(n)}_{2,k}} 2 \tilde{x}_k c_k^* - {\kappa^{(n)}_{3,k}} \tilde{G}_{k,k}|c_k|^2 \mright\} \label{eq:F_gen} \\
    \tilde{I}^{(n)}_{k,\ell}(c_k,c_\ell) &:= \lambda^{(n)}_{k,\ell} 
    \mleft( \tilde{J}_{k,\ell}(c_k, c_\ell) +  \tilde{J}_{\ell,k}(c_\ell, c_k) \mright)\\
    \tilde{J}_{k,l}(c_k,c_{\ell}) &:= -\frac{1}{\sigma^2} {\text{Re}} \mleft\{  \tilde{G}_{k,\ell} c_\ell c_k^*  \mright\}
    .
\end{align*}
\endgroup
Together with the \ac{NBP} weights, the set of optimizable parameters for the $\text{GFG}(\vek{P})$ algorithm is 
\begingroup\abovedisplayskip=4pt\belowdisplayskip=0pt
\begin{align*}
    \mathcal{P}_{\text{GFG}}(L_\text{p}) :=  \Big\{ &\vek{p} \in \mathbb{R}^{L_\text{p}}, \kappa_{i,k}^{(n)}, \lambda_{k,\ell}^{(n)}, \, k=1,\ldots,K\!+\!L,\nonumber \\ 
    &k-\ell = 1,\ldots,L, \, i=1,2,3 \Big\}  \cup  \mathcal{P}_\text{UFG}(L_\text{p}).
\end{align*}
\endgroup

\subsection{Parameter Optimization}\label{subsec:opt}
We optimize the parameter sets $\mathcal{P}_\text{FFG}$, $\mathcal{P}_\text{UFG}$ and $\mathcal{P}_\text{GFG}$ of the proposed symbol detectors towards an objective function in an end-to-end manner.
Since the factor graph-based algorithms embody \acp{DNN}, 
we rely on a rich pool of advanced optimization and training methods
developed for feed-forward \acp{DNN} in the last years.
We employ the Adam algorithm~\cite{kingma_adam_2015} as a stochastic gradient descent optimizer.
The gradients can be computed using backpropagation \cite{rumelhart_learning_1986}, which is a standard method for \acp{DNN}. All weights are initialized with~$\num{1.0}$ and the initial filter taps $\vek{p}$ of the GFG algorithm are independently sampled from a standard normal distribution.

We optimize the parametrization towards a maximum achievable rate between the channel input and the detector output.
Many practical transmission systems use \ac{BICM}, which decouples the 
symbol detection from a binary soft-decision \ac{FEC}~\cite{Fabregas_foundations_2008}.
In \ac{BICM}, the symbol detector soft output $\hat{P}(c_k|\vek{y})$ is converted by a \ac{BMD} to binary soft information 
\begin{equation*}
    \hat{P}(b_i(c_k) = \text{b} | \vek{y}) = 
    \sum\limits_{\text{m} \in \mathcal{M}_i^{(\text{b})}} \hat{P}(c_k = \text{m} | \vek{y}), \quad \text{b} \in \{ 0,1 \}
\end{equation*}
with
$\mathcal{M}_i^{(\text{b})} := \mleft \{ \text{m} \in \mathcal{M} : b_{i}(\text{m}) = \text{b} \mright \}$.
The resulting bit-wise \acp{APP} are typically expressed in \acp{LLR}
\begin{equation*}
    L_{k,i}(\vek{y}) := \ln \mleft( \frac{\hat{P}(b_i(c_k) = 0 | \vek{y})}{\hat{P}(b_i(c_k) = 1 | \vek{y})} \mright).
\end{equation*}
After interleaving, the \acp{LLR} are fed to a bit-wise soft-decision \ac{FEC}.
By interpreting the \ac{BMD} as a mismatched detector, the \ac{BMI} is an achievable information rate for \ac{BICM}~\cite{Fabregas_foundations_2008}.
The calculation of the \ac{BMI}, detailed in~\cite{alvarado_achievable_2018}, 
considers the \ac{BMD} by assuming $m$ parallel sub-channels 
transmitting on a binary basis
instead of one symbol-based channel. Assuming independent transmit bits,
the \ac{BMI} is defined as the sum of mutual informations\footnote{The mutual information is a measure between two random variables. We avoid a distinct notation for random variables as it is clear from the context.}
$I(b_i ; y)$ of $m$ unconditional bit-wise channel transmissions:
\begin{equation*}
    \text{BMI} := \sum _{i=1}^{m}\! I(b_i(c_k) ; y).
\end{equation*}
By a sample mean estimation over $D$ labeled data batches
$\mathcal{D} := \{ (\vek{c}^{(\text{d})}, \vek{y}^{(\text{d})})_{i}:
\vek{c}^{(\text{d})} \in \mathcal{M}^K,
\vek{y}^{(\text{d})} = \vek{H} \vek{c}^{(\text{d})} + \vek{w}_i , i=1,\ldots,D \}$,
a feasible approximation
\begin{align}\label{bmi_est}
    &\text{BMI} \approx \log_2 \mleft( M \mright) - \\
    &\frac{1}{D K} \sum\limits_{\mathclap{i=1}}^{m} 
    \sum\limits_{k=0}^{K-1}
    \sum\limits_{\mathclap{\qquad \quad (\vek{c}^{(\text{d})}, \vek{y}^{(\text{d})})\in \mathcal{D}}} 
    \log_2 \mleft (  \exp\mleft(-(-1)^{b_{k,i}(c_k^{(\text{d})})} L_{k,i}(\vek{y})\mright) + 1  \mright ) \nonumber
\end{align}
can be found~\cite{alvarado_achievable_2018}.
For gradient descent optimization, we maximize the objective function in~\eqref{bmi_est}.
\section{Numerical Results}\label{sec:results}
We evaluate the considered algorithms towards their detection performance.
In this paper, we show the \ac{BER} results, however,
an evaluation with respect to the \ac{BMI} shows a qualitatively similar behavior.
A linear \ac{MMSE} equalizer~\cite[Sec.~9.4]{proakis_digital_2007} with filter order $30$ as well as
the symbol-wise \ac{MAP} detector, implemented using the BCJR algorithm~\cite{bahl_optimal_1974}, serve as references.
We consider a block length of ${K=500}$ symbols and a transmission over two standard \ac{ISI} channel models: the channel \emph{Proakis~A} with an impulse response $\vek{h}_\text{A} := (0.04, -0.05,\allowbreak 0.07,\allowbreak -0.21,\allowbreak -0.5,\allowbreak 0.72,\allowbreak 0.36,\allowbreak 0.0,\allowbreak 0.21,\allowbreak 0.03, 0.07)^\text{T}$ and the channel \emph{Proakis~B}
with $\vek{h}_\text{B} := (0.407, 0.815, 0.407)^\text{T}$.

\begin{figure}[t]
\centering
    \begin{tikzpicture}[baseline, trim axis right]
    \begin{axis}[
          width=\linewidth, %
          height=0.8\linewidth,
          grid=major, %
          grid style={gray!30}, %
          xlabel= $E_\text{b}/N_0$ (dB),
          ylabel= BER,
          ymode = log,
          enlarge x limits=false,
          enlarge y limits=false,
          xmin = 0,
          ymax = 0.3,
          legend style={at={(.05,0.05)},anchor=south west, font=\small},
          legend cell align={left},
        ]
        \addplot[mark=diamond,mark options={scale=2.5,solid}, color=black, line width=1.3pt] table[x=Eb/N0, y=BCJR_BER ,col sep=comma] {numerical_results/proakisB/BPSK.csv};
        
        \addplot[densely dotted, color=gray, line width=1.3pt] table[x=Eb/N0, y=MMSE_BER ,col sep=comma] {numerical_results/proakisB/BPSK.csv};
        \addplot[dashed, mark=*,mark options={scale=0.8,solid}, color=KITred, line width=1.3pt] table[x=Eb/N0, y=FGEQ_BER ,col sep=comma] {numerical_results/proakisB/BPSK.csv};
        \addplot[mark=*,mark options={scale=0.8,solid}, color=KITred, line width=1.3pt] table[x=Eb/N0, y=NBP_FGEQ ,col sep=comma] {numerical_results/proakisB/BPSK.csv};
        \addplot[dashed, mark=pentagon*,mark options={scale=1.3,solid}, color=KITpalegreen, line width=1.3pt] table[x=Eb/N0, y=UB_FGEQ_BER ,col sep=comma] {numerical_results/proakisB/BPSK.csv};
        \addplot[mark=pentagon*, KITpalegreen, line width=1.3pt , mark options={scale=1.3,solid}] table[x=Eb/N0, y=NBP_UB_FGEQ*_BER ,col sep=comma] {numerical_results/proakisB/BPSK.csv};
        \addplot[dashed, mark=square*, KITblue, line width=1.3pt , mark options={scale=1.3,solid}] table[x=Eb/N0, y=GFG_P7*_BER ,col sep=comma] {numerical_results/proakisB/BPSK.csv};
        \addplot[KITblue, mark=square*, line width=1.3pt , mark options={scale=1.3,solid}] table[x=Eb/N0, y=NBP_GFG_P7*_BER ,col sep=comma] {numerical_results/proakisB/BPSK.csv};

        \legend{MAP \\MMSE \\ FFG \\ FFG (NBP) \\ UFG \\ UFG (NBP) \\ $\text{GFG}(\vek{P^\star})$ \\ $\text{GFG}(\vek{P^\star})$ (NBP) \\}
    \end{axis}
\end{tikzpicture}

    \caption{\ac{BER} over $\ebno$ on the Proakis~B channel with BPSK signaling for different symbol detection schemes.}
    \label{fig:proakisB_BPSK_BER}
    \vspace{-8pt}
\end{figure}
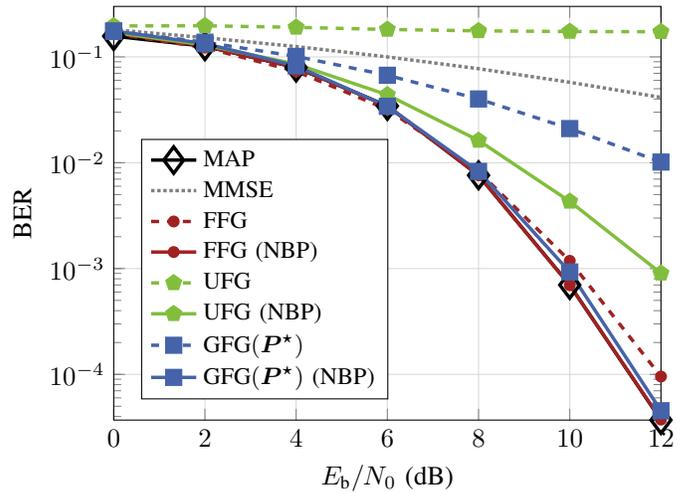
Figure~\ref{fig:proakisB_BPSK_BER} shows the \ac{BER} performance of the considered algorithms as a function of $\ebno := (m \sigma^2)^{-1}$ 
for a \ac{BPSK} transmission over the Proakis~B channel.
While the performance of the FFG detector is already close to optimal \ac{MAP} detection, the UFG algorithm performs very poorly. Notably, the \ac{BER} does not decrease for an increasing $\ebno$. We apply \ac{NBP} for both detection schemes and optimize $\mathcal{P}_\text{FFG}$ and $\mathcal{P}_\text{UFG}$ for ${\ebno = 10~\text{dB}}$.
Both algorithms show a performance improvement over the complete $\ebno$ range. The optimized FFG detector approaches optimal \ac{MAP} performance and the UFG algorithm shows a significant \ac{BER} improvement:  the application of \ac{NBP} reduces the \ac{BER} by a factor of more than $100$ for ${\ebno = 12~\text{dB}}$ .

We further examine the impact of the observation model on the UFG algorithm. %
Therefore, we generalize the preprocessing of the ${\text{UFG} = \text{GFG}(\vek{P}=\vek{H}^{\textrm H})}$ algorithm to a generic preprocessor $\vek{P^\star}$ with $L_\text{p}=7$, which we optimize with respect to the \ac{BMI} for ${\ebno = 10~\text{dB}}$.
The $\text{GFG}(\vek{P^\star})$ equalizer reaches a \ac{BER} of $10^{-2}$ at ${\ebno = 12~\text{dB}}$, 
thereby outperforming the \ac{MMSE} reference.
By applying \ac{NBP} on this modified factor graph, we achieve a detection performance which is close to optimum.

\begin{figure}[t]
    \centering
    \begin{tikzpicture}
    \begin{axis}[
          width=\linewidth, %
          height=0.8\linewidth,
          grid=major, %
          grid style={gray!30}, %
          xlabel= $E_\text{b}/N_0$ (dB),
          ylabel= BER,
          ymode = log,
          enlarge x limits=false,
          enlarge y limits=false,
          xmin = 0,
          xmax = 16,
          ymax = 0.3,
          ymin = 0.0000038,
          legend style={at={(.03,0.03)},anchor=south west, font=\small},
          legend cell align={left},
        ]
        \addplot[mark=diamond,mark options={scale=2.5,solid}, color=black, line width=1.3pt] table[x=Eb/N0, y=BCJR_BER ,col sep=comma] {numerical_results/proakisA/BPSK.csv};
        \addplot[densely dotted, color=gray, line width=1.3pt] table[x=Eb/N0, y=MMSE_BER ,col sep=comma] {numerical_results/proakisA/BPSK.csv};
        \addplot[dashed, mark=*,mark options={scale=0.8,solid}, color=KITred, line width=1.3pt] table[x=Eb/N0, y=FGEQ_BER ,col sep=comma] {numerical_results/proakisA/BPSK.csv};
        \addplot[mark=*,mark options={scale=0.8,solid}, color=KITred, line width=1.3pt] table[x=Eb/N0, y=NBP_FGEQ_BER ,col sep=comma] {numerical_results/proakisA/BPSK.csv};
        \addplot[dashed, mark=pentagon*,mark options={scale=1.3,solid}, color=KITpalegreen, line width=1.3pt] table[x=Eb/N0, y=UB_FGEQ_BER ,col sep=comma] {numerical_results/proakisA/BPSK.csv};

        \addplot[mark=pentagon*, KITpalegreen, line width=1.3pt , mark options={scale=1.3,solid}] table[x=Eb/N0, y=NBP_UB_FGEQ*_BER ,col sep=comma] {numerical_results/proakisA/16QAM.csv};
        \addplot[densely dotted, color=gray, line width=1.3pt] table[x=Eb/N0, y=MMSE_BER ,col sep=comma] {numerical_results/proakisA/16QAM.csv};
        \addplot[dashed, mark=pentagon*,mark options={scale=1.3,solid}, color=KITpalegreen, line width=1.3pt] table[x=Eb/N0, y=UB_FGEQ_BER ,col sep=comma] {numerical_results/proakisA/16QAM.csv};
    
    \node[coordinate] (A) at (5,0.01) {};
    \draw [] (A) ellipse [x radius=1.5em, y radius=0.5em, rotate=45 ];
    \node[below left=1.2em of A.south west, anchor=north east](BPSK){BPSK};
    
    \node[coordinate] (B) at (7,0.035) {};
    \draw [] (B) ellipse [x radius=1.3em, y radius=0.5em, rotate=45 ];
    \node[above right=0.7em of B.north east, anchor=south west](qam16){16-QAM};

        \legend{MAP \\
        MMSE \\
        FFG \\
        FFG (NBP) \\
        UFG \\
        UFG (NBP) \\}
        
    \end{axis}
\end{tikzpicture}

\caption{Hard-decision performance of different symbol detectors for a \ac{BPSK} and a 16-\ac{QAM} transmission over the Proakis~A channel.}
\label{fig:proakisA}
\vspace{-8pt}
\end{figure}
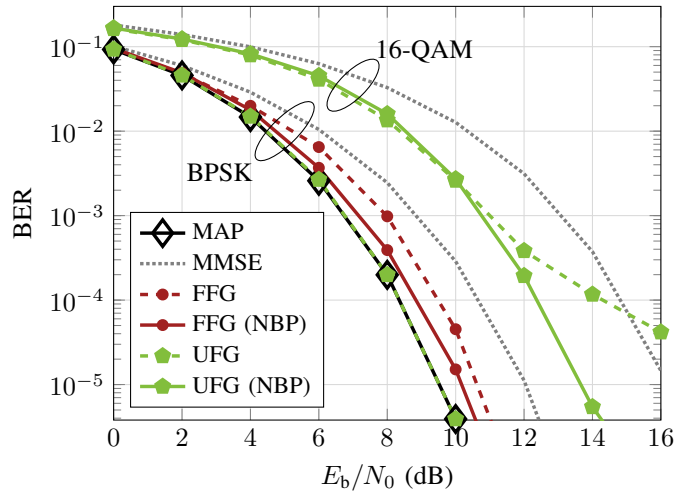

The detection performance on the Proakis~A channel is evaluated in Fig.~\ref{fig:proakisA}.
For \ac{BPSK} signalling, the UFG algorithm shows already a quasi-optimum performance, whereas the FFG detector has a loss of ${1~\text{dB}}$ at ${\ebno = 10^{-5}}$.
The application of \ac{NBP} improves the detection performance, but cannot entirely close the gap to the \ac{MAP} performance.
For 16-\ac{QAM}, the FFG and the BCJR algorithm become computationally infeasible. The UFG algorithm outperforms the \ac{MMSE} algorithm in the low $\ebno$ regime but runs into an error floor. Optimizing the UFG detector for ${\ebno = 14~\text{dB}}$ compensates this behavior and generalizes well over the complete $\ebno$ range. The GFG algorithm with $L_\text{p} \geq L=10$ yields similar results (not shown here to keep the figure simple).
\section{Conclusion}\label{sec:conclusion}
We considered the application and neural enhancement of message passing on cyclic factor graphs for symbol detection on \acl{ISI} channels.
Performance and computational complexity of the algorithms strongly depend on the underlying factor graph, which can be actively varied by a change in the observation model.
We proposed simple but effective generalizations of the factor graph, 
as well as \ac{NBP}---as an enhanced message passing algorithm---in order to mitigate the effect of cycles in the graphs.
The methods are only marginally increasing the detection complexity.
Our numerical evaluations demonstrate the effectiveness of the proposed methods by reaching near-\ac{MAP} performance for two exemplary channel models.
Especially for high-order constellations and static channels with large memory, the proposed novel GFG algorithm is a promising low-complexity alternative to the BCJR.

\end{document}